\documentclass[journal]{IEEEtran}
\usepackage{ifpdf}
\usepackage{cite}

\def\flagarxiv{x}

\ifCLASSINFOpdf
  \usepackage[pdftex]{graphicx}
  \graphicspath{{images/}}
  \DeclareGraphicsExtensions{.pdf,.jpeg,.png}
\fi

\usepackage{amsmath}
\usepackage{algorithmic}
\usepackage{array}

\ifCLASSOPTIONcompsoc
  \usepackage[caption=false,font=normalsize,labelfont=sf,textfont=sf]{subfig}
\else
  \usepackage[caption=false,font=footnotesize]{subfig}
\fi

\usepackage{fixltx2e}
\usepackage{stfloats}

\ifCLASSOPTIONcaptionsoff
  \usepackage[nomarkers]{endfloat}
 \let\MYoriglatexcaption\caption
 \renewcommand{\caption}[2][\relax]{\MYoriglatexcaption[#2]{#2}}
\fi
\usepackage{url}

\usepackage[utf8]{inputenc}
\usepackage{graphicx}
\usepackage{pgfplots}
\usepackage{tikz}
\usepackage{tikzscale}
\usepackage{tikz-uml}

\usepackage{adjustbox}

\usetikzlibrary{calc,positioning,shapes.multipart,shapes,shapes.geometric,arrows,fit,shadings,backgrounds,patterns,matrix}
\usetikzlibrary{arrows.meta, chains, shapes.multipart}

\tikzstyle{arrow} = [thick,->,>=stealth]
\tikzstyle{dashedarrow} = [thick,->,>=stealth, dashed]

\usepackage{listings}
\usepackage[acronym, nomain]{glossaries}

\usepackage{booktabs}
\usepackage{fontawesome}
\usepackage{xspace}
\usepackage[inline]{enumitem}
\usepackage{pgfplotstable}
\usepackage{colortbl}
\usepackage{multirow}
\PassOptionsToPackage{hyphens}{url}
\usepackage[hidelinks]{hyperref}
\usepackage{amsmath}
\usepackage{amssymb}
\usepackage{pifont}
\usepackage{makecell}

\usepackage[hyphens]{url}
\usepackage{hyperref}
\usepackage[hyphenbreaks]{breakurl}

\usepackage[color=blue!30]{todonotes}

\usepackage{tabularx}

\newcommand{\ie}{i.e.,\xspace}
\newcommand{\eg}{e.g.,\xspace}

\newcommand{\class}[1]{#1\xspace}
\newcommand{\rel}[1]{#1\xspace}

\newacronym{securitycapability}{SC}{Security Capability}

\newacronym{securitycapabilitymodel}{SCM}{Security Capability Model}
\newcommand{\scm}{\gls{securitycapabilitymodel}\xspace}

\newacronym{networksecurityfunction}{NSF}{Network Security Function}
\newcommand{\nsf}{\gls{networksecurityfunction}\xspace}

\newacronym{securitycontrol}{SC}{Security Control}
\newcommand{\seccon}{\gls{securitycontrol}\xspace}

\newacronym{refinementengine}{RefEng}{Refinement Engine}
\newcommand{\refeng}{\gls{refinementengine}\xspace}

\newacronym{remediationmodule}{ReM}{Remediation Module}
\newcommand{\remeng}{\gls{remediationmodule}\xspace}

\newacronym{cataloguemanager}{CaM}{Catalogue Manager}
\newcommand{\cam}{\gls{cataloguemanager}\xspace}

\newacronym{enforcer}{Enf}{Enforcer}

\newacronym{informationmodel}{IM}{Information Model}
\newcommand{\im}{\gls{informationmodel}\xspace}

\newacronym{datamodel}{DM}{Data Model}

\newacronym{middlelevelpolicy}{MLP}{Middle-Level Policy}
\newcommand{\mlp}{\gls{middlelevelpolicy}\xspace}

\newacronym{highlevelpolicy}{HLP}{High-Level Policy}
\newcommand{\hlp}{\gls{highlevelpolicy}\xspace}

\newacronym{lowlevelconfiguration}{LLC}{Low-Level Configuration}
\newcommand{\llc}{\gls{lowlevelconfiguration}\xspace}

\newboolean{arxiv}
\ifdefined\flagarxiv
\setboolean{arxiv}{true}
\else
\setboolean{arxiv}{false}
\fi

\newcommand{\fcSecCap}{SecurityCapability}

\newcommand{\fcEventCap}{EventCapability}
\newcommand{\fcDefActCap}{DefaultActionCapability}
\newcommand{\fcActCap}{ActionCapability}
\newcommand{\fcEvalCap}{EvaluationCapability}
\newcommand{\fcCondCap}{ConditionCapability}
\newcommand{\fcResStrCap}{ResolutionStrategyCapability}

\ifthenelse{\boolean{arxiv}}
{
\newcommand{\changed}[1]{{\color{black}#1}}
}
{
\newcommand{\changed}[1]{{\color{blue}#1}}
}

\begin{document}
%
\title{A Formal Model of Security Controls' Capabilities and Its Applications to Policy Refinement and Incident Management}

\author{Cataldo Basile, Gabriele Gatti, Francesco Settanni
\thanks{Cataldo Basile, Gabriele Gatti, and Francesco Settanni are with the Dipartimento di Automatica e Informatica, Politecnico di Torino, Torino, Italy, e-mail: {\it \{name.surname\}@polito.it}}
}

%

\markboth{IEEE/ACM TRANSACTIONS ON NETWORKING,~Vol.~XX, No.~YY, Xxxxx~2024}%
{Canavese \MakeLowercase{\textit{et al.}}: Asset identification}

\maketitle

\begin{abstract}
Enforcing security requirements in networked information systems relies on security controls to mitigate the risks from increasingly dangerous threats.
Configuring security controls is challenging; even nowadays, administrators must perform it without adequate tool support. 
Hence, this process is plagued by errors that translate to insecure postures, security incidents, and a lack of promptness in answering threats.

This paper presents the \scm, a formal model that abstracts the features that security controls offer for enforcing security policies, which includes an Information Model that depicts the basic concepts related to rules (i.e., conditions, actions, events) and policies (i.e., conditions' evaluation, resolution strategies, default actions), and a Data Model that covers the capabilities needed to describe different types of filtering and channel protection controls.
Following state-of-the-art design patterns, the model allows for generating abstract versions of the security controls' languages and a model-driven approach for translating abstract policies into device-specific configuration settings.
By validating its effectiveness in real-world scenarios, we show that \scm enables the automation of different and complex security tasks, i.e., accurate and granular security control comparison, policy refinement, and incident response. 
Lastly, we present opportunities for extensions and integration with other frameworks and models.
 
\glsresetall
\end{abstract}


\IEEEpeerreviewmaketitle



\ifthenelse{\boolean{arxiv}}{
  \begin{tikzpicture}[remember picture,overlay]
    \node[inner sep=5pt,right, text width  = 1.06\textwidth, fill=yellow,draw,line width = 1pt ,fill=yellow!20,execute at begin node=\setlength{\baselineskip}{3pt}] at ($(current page.north west) + (1cm,-27cm)$) 
    { \footnotesize \color{red} This work has been submitted to the IEEE/ACM for possible publication. Copyright may be transferred without notice, after which this version may no longer be accessible.};
  \end{tikzpicture}%
}
{}



\section{Introduction}\label{sec:intro}
\seccon{}s, such as firewalls and VPN gateways, are crucial to mitigate the cybersecurity risks that threaten the organizations' networks.

Recent networking technologies that deviate from on-premise infrastructure management, like cloud computing, have increasingly hidden these devices from users and administrators; nonetheless, they still need to be configured.
However, the configuration of \seccon{} has been historically associated with human errors \cite{Wool10cheese}. They are complex to configure, and the complexity grows with the size and heterogeneity of the networks, the number of entities to consider (e.g., hosts, services, users), and the required rules in their configuration files \cite{al-shaer-taxonomy}. 
Unfortunately, the tools to help users reduce human errors and support the management of \seccon{}s configurations never became established due to their limitations and lack of generalization~\cite{bringhenti2023, firewall-usability-litrev}. 

Methods for automatic security configuration using refinement or anomaly analysis techniques have been investigated in the last years, but they have had a minimal impact on the practice \cite{basile2019ton,bringhenti23automated}.
Analogously, other activities in complex distributed systems that rely on configuring \seccon{}s, such as incident management (e.g. reacting to attacks to limit the negative consequences) and risk mitigation (e.g. configuring \seccon{}s to prevent attacks), require promptness and human expertise. The former can be appropriately achieved by delegating the tasks to computer processes, while the latter involves advanced modelling of security experts. Solutions satisfying these requirements may be faster and, if adequately validated, less prone to errors, yet they still need to be achieved \cite{Chouhan2023IntrusionRS, INAYAT201653, security-orchestration-review}.

Significant efforts are needed to investigate novel techniques to bridge the gaps in the current state of the art. 
This research work is based on the hypothesis that the limitations in the current practices could also originate from the lack of formal models of \seccon{}s, which may capture the hidden semantics that allow human beings to operate. 
Hence, this paper aims to answer a crucial question: \textit{Can a formal model of \seccon{}s improve the network security landscape by mitigating some of the current issues plaguing it?}

To this purpose, this paper introduces a formal model of security control capabilities, the \scm, which represents the configurations these devices can accept based on their specific languages. 
In other words, the model captures what \seccon{}s can do to enforce security policies.

We have analyzed several open-source \seccon{}s to highlight the generic concepts that characterize the configuration languages, regardless of individual controls.
Then, we have modelled selected \seccon{}s, formally describing their capabilities, by 
\begin{enumerate*}
\item characterizing their conditions by the data type they accept and the logical operators they support, and 
\item enumerating all the actions, based on the effect they produce on the resources to which they are applied, by
\item modelling how these can be used to form valid policies. 
\end{enumerate*}

These abstractions formed the initial definition of the \scm, which allows performing several tasks, e.g., comparing different \seccon{}s to determine when they can enforce the same actions and when they have conditions in common. 
The model supports informed decisions when selecting a \seccon that could implement a specific policy, comparing two products to determine the best one to use, and migrating from a control to a different one from another vendor. \changed{It is worth noting that our objective is to provide a versatile model independent of both the application scenario and the attack model, which only determine the \seccon{}s that will be employed.}

The initial definition has been enriched to support other security tasks: a set of association classes added to the model determines how abstract capabilities can be converted into low-level configuration settings understandable by a target \seccon. This feature enables translations between different levels of abstraction (i.e., policy refinement), as well as among different \seccon{}s (two firewalls from distinct vendors).

The practical outcomes of this research include the \scm\footnote{This model's preliminary version has already been presented in a previous publication \cite{basile-model-capabilities}.}, a UML model from which XML representations can be obtained, a Catalogue with different \seccon{}s from various categories, a set of web services for managing models and querying catalogues, and a policy translation engine.

\changed{The usefulness of the \scm lies in its ability to decompose security controls into their features, providing an interface for ensuring vendor-neutral security policy management, and support daily tasks, like helping administrators when (manually) translating high-level security requirements, e.g., stemming from regulations or internal security policies, into actual configurations or to lock and isolate threats when reacting to security incidents.
With this regard, the \scm has been validated with three relevant real-world scenarios:
}
\begin{itemize}
\item \textit{Security management tasks}, the first scenario involved the evaluation of how the formal descriptions in the security capability model can reduce vendor lock-in;

\item \textit{Automatic configuration of software networks}, the second scenario dealt with developing a tool for refining security policies. The tool utilizes capabilities to identify suitable \seccon{}s for policy enforcement and employs a translator to generate the appropriate configurations.

\item \textit{Automated remediation}, lastly, the third scenario tested the ability to remediate security incidents with playbooks that use capabilities to determine the \seccon{}s to reconfigure or how to intervene at the network level when playbooks are not enforceable (\eg adding more \seccon{}s to mitigate risks).
\end{itemize}

Our results prove that in an era where everyone seems to want to offload their tasks to AI-based chatbots, developing formal models is still a crucial activity (yet complex and time-consuming), especially in areas like cybersecurity, where even minor uncertainty or inaccuracies cannot be tolerated.

The rest of the paper is structured as follows.
Section~\ref{sec:motivating} presents the three relevant scenarios where the security capability model has been validated.
Section~\ref{sec:capability_model} presents the \scm, the design decisions, and the features it exposes. 
Section~\ref{sec:implementation} presents the tools which rely on the \scm and that served to validate the importance of a formal model of security capabilities, as described in Section~\ref{sec:validation}.
Section~\ref{sec:related} describes past works that presented or used formal models in policy-based security.
Finally, Section~\ref{sec:conclusions} concludes and provides hints for future works.

\section{Motivating examples and requirements}\label{sec:motivating}
Three main areas of application of the \scm highlight the progress over state of the art and allow answering our research questions about the usefulness of a formal model of \seccon{}s.

\subsection{Cybersecurity market and Vendor lock-in}\label{subsec:motivating-uc1}
%
\changed{The first scenario originates from the enterprise market.
Security administrators and network designers employ various types of \seccon{}s for enforcing security in their daily activities. They may want to switch products for better performance or features, reduced costs, or because they lost trust in the original vendor. But vendors hinder this process.}

The \seccon{}s selection process has been widely studied in the literature \cite{BARNARD2000185,YEVSEYEVA20151035}. Still, these methods leave the final evaluation of the features offered by competing vendors to the experts \cite{selection-human-administrators} who have to guess what functionalities \seccon{}s provide and compare similar \seccon{}s. 
However, products are not easily comparable due to buzzword-driven data sheets, incremental updates and re-branding of old ideas marketed as radical improvements, and limited testing options \cite{security-software-market}.

Although not well documented in the literature, empirical observations of the enterprise security panorama showed that easy migrations between different security products are unfortunately far from reality. This generally leads to adopting same-vendor products even if better alternatives are available on the market. 
For example, changing the firewall vendor requires time-consuming translation. This translation may be assisted by migration software\footnote{\url{https://live.paloaltonetworks.com/t5/expedition/ct-p/migration_tool}\\
\url{https://www.fortinet.com/products/next-generation-firewall/forticonverter}} 
but requires manual validation; hence, it is still mostly human-dependant and prone to misconfiguration.


Therefore, our research objectives include verifying that a formal representation of the \seccon{}s using predefined security capabilities allows for
\begin{itemize}
    \item ($\mathrm{RO}_1$) a simple evaluation and comparison of the functions covered by commercial products;
    \item ($\mathrm{RO}_2$) mapping abstract policies into to ready-to-use low-level settings exported in the \seccon{}-specific language. 
\end{itemize} 
The scenario validation will be presented in Section~\ref{sec:validation:uc1}.


\subsection{Policy Refinement}\label{subsec:motivating-uc2}
Enterprise security requirements are usually maintained as abstract statements. 
However, statements can only be enforced after a security policy refinement process \cite{iperial-college-policy-refinement}.
First, one has to determine the resources, \ie the \seccon{}s, that may be used to enforce the desired requirements. Then, requirements are translated into the low-level settings for the identified \seccon{}s.

\changed{Refinement is a complex problem that security architects face regularly, and complexity increases with the size and heterogeneity of the target infrastructure.}
Currently, this process is mainly conducted by network and security administrators, a human-driven approach that has been linked to misconfigurations and conflicts \cite{al-shaer-management,al-shaer-taxonomy}. Automatic policy refinement has been proposed in the literature; however, previous works focus on vertical areas (\ie access control\cite{cheminod2019comprehensive}, channel protection \cite{valenzaAutoChannelProt}), rather than analyzing the problem from the perspective of what the generic \seccon{}s can do and the entities they interact with.
Hence, automating the refinement in real contexts is still an open problem.

Refinement could benefit from a formal model of capabilities that can describe precisely how individual \seccon{}s work. 
Indeed, the generic concepts related to configurations, such as conditions and actions, are common to all \seccon{}s. 
Different controls share several configuration parameters, data types and options (\eg IP addresses are used to filter traffic by all firewalls). Still, they have their peculiarities (\eg conditions on protocol states are typically control-specific) and configuration languages.
For this scenario, our research objectives include:
\begin{itemize}
    \item ($\mathrm{RO}_3$) accurately modelling the context, the general concepts and the individual elements that constitute the device configurations process, including allowing the selection with high precision the \seccon{}s that can be used to enforce security policies;
    \item ($\mathrm{RO}_4$) permitting the refinement of high-level security policies into configuration settings for selected \seccon{}s.
\end{itemize} 


The scenario validation will be presented in Section~\ref{sec:validation:uc2}.

\subsection{Incident response}\label{subsec:motivating-uc3}
 
\changed{Incident response teams, in their day-to-day activities, must ensure that systems are secured and return to their operational state after incidents but have to face limitations from the operational practice \cite{sok, soccc, soar-playbooks}.}

Many solutions have been proposed and designed to address this scenario over the years \cite{interop-between-cyber-systems, soarandai}, which promise optimal functionality and seamless operations and are all underscored by continuous enhancements over the others.
Solutions include SOAR, ``solutions that combine incident response, orchestration and automation, and threat intelligence platform management capabilities in a single solution'' among many other hybrids which have been given birth over the years, such as XDR \cite{gartner-soar-website, soar-sortof-survey, soarandai}.

These solutions may orchestrate security processes, provide remediation measures as security playbooks \cite{playbooks}, and interact with the available \seccon{}s (virtual and physical). In most cases, they underwent real-world validation and on-the-ground experience from security operators.
Unfortunately, as of now, these solutions lack effective integration with \seccon{}s to be considered capable of remediating or mitigating incidents \cite{soar-security-controls-interoperability,interop-between-cyber-systems}.



For this scenario, our research objectives include:
\begin{itemize}
    \item ($\mathrm{RO}_5$) defining a precise mapping between the \seccon{}s' features and the operations that be used for remediation, to mitigate risks and benefit incident management;
    \item  ($\mathrm{RO}_6$) augment cyber threat intelligence with additional context and artefacts that are useful to characterize incidents better, allowing fine-tuned reactions, sharing of supplementary data, and continuous monitoring of the appropriateness of the defensive apparatus
\end{itemize}
The scenario validation will be presented in Section~\ref{sec:validation:uc3}.


\section{The capability model}\label{sec:capability_model}
\begin{figure}[t!]
\centering
\scalebox{0.65}{

\begin{tikzpicture}

\umlemptyclass[]{NSFPolicyDetails}
\umlemptyclass[below=4ex of NSFPolicyDetails]{NSF}
\umlemptyclass[right=12ex of NSF]{SecurityCapability}
\umlemptyclass[left=5ex of NSF]{NSFCatalogue}
\umlemptyclass[below right=3ex and -10ex of NSF]{HasSecurityCapabilityDetails}
\umlemptyclass[below right=5ex and -14ex of HasSecurityCapabilityDetails]{CapabilityTranslationDetails}
\umlemptyclass[left=2ex  of CapabilityTranslationDetails]{ResolutionStrategyDetails}

\umlaggreg[]{NSF}{NSFPolicyDetails}

\umlaggreg[]{NSFCatalogue}{NSF}
\umlaggreg[geometry=|-]{NSFCatalogue}{HasSecurityCapabilityDetails}
\umlaggreg[anchor1=20, anchor2=10]{NSF}{SecurityCapability}

\umlinherit[anchor1=-10, anchor2=-20]{SecurityCapability}{NSF}

\umlassoc[geometry=|-]{NSF}{HasSecurityCapabilityDetails}
\umlassoc[geometry=-|, anchor1=east]{HasSecurityCapabilityDetails}{SecurityCapability}

\umlinherit[geometry=|-|]{ResolutionStrategyDetails}{HasSecurityCapabilityDetails}

\umlinherit[geometry=|-|]{CapabilityTranslationDetails}{HasSecurityCapabilityDetails}

\end{tikzpicture}

}

\caption{The Security Capability Information Model.}
\label{fig:details}
\end{figure}
\begin{figure}[t!]
\centering
\scalebox{0.65}{

\begin{tikzpicture}

\umlemptyclass[]{\fcSecCap}

\umlemptyclass[above left=3ex and 8ex of \fcSecCap]{\fcActCap}{}
\umlemptyclass[below=3ex of \fcActCap]{\fcDefActCap}{}
\umlemptyclass[below=2ex of \fcDefActCap]{\fcResStrCap}{}

\umlemptyclass[above right=2ex and 4ex of \fcSecCap]{\fcEvalCap}
\umlemptyclass[below=2ex of \fcEvalCap]{\fcEventCap}
\umlemptyclass[below=2ex of \fcEventCap]{\fcCondCap}

\umlinherit[geometry=-|-,anchor2=west]{\fcDefActCap}{\fcSecCap}
\umlinherit[geometry=-|-,anchor2=west]{\fcActCap}{\fcSecCap}
\umlinherit[geometry=-|-,anchor2=west]{\fcCondCap}{\fcSecCap}
\umlinherit[geometry=-|-,anchor2=east]{\fcEvalCap}{\fcSecCap}
\umlinherit[geometry=-|-,anchor2=east]{\fcEventCap}{\fcSecCap}
\umlinherit[geometry=-|-,anchor2=east]{\fcResStrCap}{\fcSecCap}
\umlaggreg[]{\fcDefActCap}{\fcActCap}

\end{tikzpicture}

}

\caption{Subclasses of SecurityCapability.}
\label{fig:6tuple}
\end{figure}

At a high level, \seccon{}s are well approximated with the IETF architecture \cite{rfc2753}, which consists of a Policy Enforcement Point that receives an object (e.g., a packet on a NIC) and asks a Policy Decision Point what to do with that object based on a security policy. In this context, the security policy is the \seccon{}'s configuration specified with its specific language.
Policies are (most often) formed by rules.
Rules are the means to specify what to enforce (i.e., the actions), where (the objects to which these actions need to be enforced, i.e., the \textit{conditions}), and when (the \textit{events}). When a rule includes more than one, conditions are organised in logical formulas (\textit{condition clause}) that allow determining when the rule matches the object.
To ensure coherence, \seccon{}s support methods to determine what to enforce when no rules apply (\textit{default actions}) {\changed{or when multiple rules match the object (\textit{resolution strategies}).}

The \scm captures these relations through an \im, presented in Fig.~\ref{fig:details}.
The \class{NSF} and \class{SecurityCapability} classes abstract \nsf{}s and generic Security Capabilities. 
Particularly, the \im implements the ``decorator pattern'' \cite{Gamma2010}.
The \class{HasSecurityCapabilityDetails} class implements the association class used to ``decorate'' the security capabilities when they are associated with an \nsf, and will be presented more in detail in Section~\ref{sec:modelling_translation}.

The aggregation of \rel{SecurityCapability} makes \class{NSF} instances containers of capabilities. Since \class{SecurityCapability} inherits from \class{NSF}, this pattern allows the inclusion of groups of specific capabilities into new \class{NSF} instances; hence, it supports creating and merging \nsf description templates.
{To make a concrete example, the capabilities of filtering controls operating at the fourth OSI layer, like IpTables, can be obtained by extending the description of a generic packet filter with additional conditions on TCP states, MAC, bandwidth,~etc.}

The \class{SecurityCapability} class has been subclassed to describe the main concepts needed to model policies and emerged during the analysis of several \seccon{}s (see Fig.~\ref{fig:6tuple}).


Four concepts are needed to model rules.
As anticipated before, the \emph{conditions} are the features available at the \nsf to trigger the rule enforcement. Conditions are expressed with specific constraints on different data types (\eg matching an IP address or checking a regex on the HTTP MIME type allows selecting the traffic to allow).
The \emph{actions} are the operations an \nsf can perform on individual elements (\eg deny packets or encrypt flows). 
The \emph{events} allow triggering the rules' evaluation for specific classes of controls (\eg the reception of a network packet on an interface).
The \emph{condition clauses} indicates the formula for determining when a rule is activated depending on the evaluation of individual conditions, the most common ones being the Disjunctive Normal Form (DNF), i.e., the rule is activated when all the conditions are true, or the Conjunctive Normal Form (CNF), i.e., it is activated when just one condition is satisfied.

Two more concepts illustrate how to build security policies from individual rules.
The \textit{resolution strategies} describe how the \seccon behaves when multiple rules are fired. For instance, the First Matching Rule strategy applies the actions from the highest priority rule.
The \textit{default actions} determines how the \seccon behaves when no rule is fired (e.g., for implementing the ``deny all'' default behaviour). If some actions are not available at an NSF to specify default behaviour, the \class{DefaultAction} class allows aggregating the allowed actions.
 
Finally, the \class{NSFCatalogue} class is the central aggregator of \class{NSF} instances and decorations and the root element of the repositories.
More details on \scm classes and attributes are available in a technical report\footnote{\url{https://github.com/aldobas/SecurityCapabilityModelPub/tree/main/manual}}.


\subsection{Modelling the reference security controls} \label{sec:dm}

Individual capabilities are modelled with a Data Model.
For this research, we have analysed controls that filter traffic up to OSI layer 4 and layer 7, channel protection and VPN controls, namely:
\begin{itemize}   
    \item IpTables\footnote{\url{https://netfilter.org/news.html}}, a very widespread and effective filtering control available in the Linux distributions (only the filtering modules, no NAT/NAPT);
    \item XRFM\footnote{\url{https://man7.org/linux/man-pages/man8/ip-xfrm.8.html}}, the native IPsec configuration for Linux;
    \item strongswan\footnote{\url{https://www.strongswan.org/}}, an advanced IPsec/IKE implementation;
    \item squid\footnote{\url{http://www.squid-cache.org/}}, a web caching proxy with an application layer filtering module.
\end{itemize}

Moreover, the Data Model includes controls that abstract the features of theoretical \seccon{}s. Fig.~\ref{lst:gpf} presents an XML description of a generic packet filter, which serves to specify the allowed or denied connections using five-tuples composed of source and destination IPs and ports, and protocol type \cite{al-shaer}. 
Fig.~\ref{lst:gpf} also presents the definition of one of the condition capabilities, the \class{IpSourceAddressConditionCapability} and its data type (i.e., IpSourceAddressType), defined within an XMLSchema.

\begin{figure}[tb]
\begin{adjustbox}{max width = .49\textwidth}
\begin{lstlisting}[basicstyle=\scriptsize\ttfamily] 
<nSF id="genericPacketFilter">
   <securityCapability ref="IpProtocolTypeConditionCapability"/>
   <securityCapability ref="IpSourceAddressConditionCapability"/>
   <securityCapability ref="IpDestinationAddressConditionCapability"/>
   <securityCapability ref="SourcePortConditionCapability"/>
   <securityCapability ref="DestinationPortConditionCapability"/>
   <securityCapability ref="AcceptActionCapability"/>
   <securityCapability ref="RejectActionCapability"/>
   <securityCapability ref="DefaultActionCapabilitySpec"/>
</nSF>
<securityCapability id="IpSourceAddressConditionCapability"
   xsi:type="IpSourceAddressType"/>
\end{lstlisting}
\end{adjustbox}
\caption{Specification of the Generic packet Filter}
\label{lst:gpf}
\end{figure}

The definition of the IpTables \nsf can be obtained by extending the definition of the generic packet filter with its resolution strategy (i.e., the \class{FMRResolutionStrategyCapabilitySpec}), the additional conditions (187 ConditionCapability classes) and actions (140 ActionCapability classes) it supports.

The \nsf Catalogue\footnote{The Catalogue and the \scm Information and Data Models can be found at \url{https://github.com/aldobas/SecurityCapabilityModelPub}} describes seven \seccon{}s. The Data Model contains one single instance of capabilities for events, and Resolution Strategy (the FMR), two evaluations (CNF and DNF), one default action (the class that indicates that all the actions are suitable as default); considering all the \seccon{}s, it contains 270 condition and 167 action capabilities.

\subsection{Deriving the abstract language for a NSF}

The capabilities describe what \seccon{}s can do, which also correspond to the features that can be enabled with their configuration languages. 
Hence, capabilities can be associated with abstract language constructs. 
Starting from \class{NSF} instances, an automatic language generation process can derive their \nsf Abstract Languages. The \nsf Abstract Language can be used to specify the configuration for that specific \nsf, named \mlp.
\mlp{}s use a syntax that is independent of the \seccon{} for which they have been derived, but since they are validated against the \nsf Abstract Language, it is ensured that they only contain instructions for capabilities owned by the \seccon{}.

Fig.~\ref{lst:language} presents the abstract language of the generic packet filter in Fig.~\ref{lst:gpf}. 
The \class{Policy} class aggregates \class{Rules}. \class{Rule} instances only allow using conditions and actions taken from the \nsf specification; they are associated with the proper data types. Moreover, rules include the necessary fields for supporting the resolution strategy and attributes helpful for specification purposes (e.g., descriptions and labels).

\begin{figure}[tb]
\begin{adjustbox}{max width = .485\textwidth}
\begin{lstlisting}[basicstyle=\scriptsize\ttfamily] 
<xs:complexType name="Policy">
<xs:sequence>
   <xs:attribute name="nsfName" type="xs:string"/>
   <xs:element maxOccurs="1" minOccurs="0" 
        name="defaultActionCapabilitySpec" 
        type="DefaultActionCapability"/>
   <xs:element maxOccurs="unbounded" minOccurs="0" 
        name="rule" type="Rule"/>
</xs:sequence>
</xs:complexType>
<xs:complexType name="Rule">
    <xs:choice maxOccurs="unbounded">
        <xs:element name="ruleDescription" type="xs:string"/>
        <xs:element name="label" type="xs:string"/>
        <xs:element name="externalData" type="ExternalData"/>
        <xs:element name="defaultActionCapabilitySpec" 
                    type="DefaultActionCapabilitySpec"/>
        <xs:element name="ipSourceAddressConditionCapability" 
                    type="IpSourceAddressConditionCapability"/>
        <xs:element name="ipDestinationAddressConditionCapability" 
                    type="IpDestinationAddressConditionCapability"/>
        <xs:element name="sourcePortConditionCapability" 
                    type="SourcePortConditionCapability"/>
        <xs:element name="destinationPortConditionCapability" 
                    type="DestinationPortConditionCapability"/>
        <xs:element name="ipProtocolTypeConditionCapability" 
                    type="IpProtocolTypeConditionCapability"/>
        <xs:element name="acceptActionCapability" 
                    type="AcceptActionCapability"/>
        <xs:element name="rejectActionCapability" 
                    type="RejectActionCapability"/>
    </xs:choice>
    <xs:attribute name="id" type="xs:string"/>
    <xs:attribute name="ruleType" type="xs:string"/>
</xs:complexType>
<xs:complexType name="ExternalData">
    <xs:simpleContent>
        <xs:extension base="xs:string">
            <xs:attribute name="type" type="xs:string"/>
        </xs:extension>
    </xs:simpleContent>
</xs:complexType>
\end{lstlisting}
\end{adjustbox}
\caption{Abstract language of the Generic Packet Filter}
\label{lst:language}
\end{figure}

\subsection{Model-driven approach for translation}\label{sec:modelling_translation}
    
\mlp{}s are not directly usable to configure \seccon{}s; hence, a further translation into the appropriate low-level settings is needed. An essential requirement in designing this translation process was to avoid constantly updating the translator code.

The decorator pattern helped us reach our goal, as it allows providing additional information when capabilities are ``attached'' to the decorated objects.
It includes the \rel{HasSecurityCapabilityDetails} association class, which in the implementation is associated with both \class{NSF} and \class{SecurityCapability}.
This class can be used to map abstract capabilities to device-specific settings at the model level, permitting the development of a model-driven translator that does not need \nsf-specific code to perform the translations. Indeed, translation directives are available as association classes. 

Different types of ``details'' (see Fig.~\ref{fig:details}) were needed in the \scm.
\class{NSFPolicyDetails} aggregates instructions for generating valid policies for the target \nsf and is described in Section~\ref{sec:nsfpolicydetails}).
\class{CapabilityTranslationDetails} specifies how to translate abstract
capabilities into low-level settings and is discussed in Section~\ref{sec:CapabilityTranslationDetails}
\class{ResolutionStrategyDetails} specifies the external data the Resolution Strategy uses, which are sketched in Section~\ref{sec:ResolutionStrategyDetails}.

\subsection{NSFPolicyDetails}\label{sec:nsfpolicydetails}

\begin{figure}[tb]
\begin{adjustbox}{max width = .49\textwidth}
\begin{lstlisting}[basicstyle=\scriptsize\ttfamily]
<nSF id="IpTables">
  <nsfPolicyDetails>
    <ruleStart>iptables</ruleStart>
    <policyAttribute>
      <attributeName>targetRuleSet</attributeName>
    </policyAttribute>
    <defaultSecurityCapability>appendRuleActionCapability
    </defaultSecurityCapability>
  </nsfPolicyDetails>
  ...
</nSF>
\end{lstlisting}
\end{adjustbox}
\caption{Policy details for IpTables}
\label{lst:iptablesdetails}
\end{figure}

The \class{NSFPolicyDetails} class conveys information on how to translate policy-related information for a target \nsf; it has been implemented as an inner attribute of the \class{NSF} class. It reports:
\begin{itemize}
    \item strings and formats needed to generate valid rules and policies (with the attributes \texttt{ruleStart}, \texttt{ruleEnd}, \texttt{policyTrailer}, and \texttt{policyEncoding});
    \item the capabilities that need to mandatorily be present for a valid policy, which may vary depending on the \seccon{} (\texttt{defaultSecurityCapability});
    \item other parameters to be given to the Translator for a correct generation that cannot be obtained otherwise (\texttt{policyAttribute}).
\end{itemize}
Fig.~\ref{lst:iptablesdetails} presents the \nsf details for Iptables. Rules must start with the `iptables' prefix; every IpTables rule must contain an option that indicates the policy chain where the rules need to be inserted (e.g., the \texttt{-A} option for appending them). The chain name where rules will be appended (e.g., INPUT, OUTPUT, FORWARD) is passed as an attribute.

\subsection{CapabilityTranslationDetails}
\label{sec:CapabilityTranslationDetails}

The \class{CapabilityTranslationDetails} class conveys data needed to translate individual capabilities to form correct \nsf-specific rules.
In other words, the attribute values reported with this class specify the low-level policy rule semantics and syntax and are crucial for the model-driven translation.
It includes the keywords to invoke the capability (\texttt{realCommandName}) and additional attributes to adjust the command name when Boolean operators are used, e.g., to negate an individual condition (\texttt{commandNameCondition}), and how this keyword connects to the values (e.g., to put an equal after the condition name and a semicolumn to end it with), as in the example below for source port conditions in IpTables:

\begin{figure}[tb]
\begin{adjustbox}{max width = .49\textwidth}
\begin{lstlisting}[basicstyle=\scriptsize\ttfamily]
<securityCapability ref="SourcePortConditionCapability" />
<commandName>
  <realCommandName>--sport</realCommandName>
</commandName>
<commandName>
  <realCommandName>! --sport</realCommandName>
  <commandNameCondition>
    <attributeName>operation</attributeName>
    <attributeValue>NOT_EQUAL_TO</attributeValue>
  </commandNameCondition>
</commandName>
\end{lstlisting}
\end{adjustbox}
\caption{Source port connection}
\label{lst:source_port}
\end{figure}

Such class also reports the way to parse the values from the \mlp and merge them into a valid statement (\texttt{bodyConcatenator}), how to provide more individual values (e.g. \lstinline|port=80,81,82|), build ranges (e.g., \lstinline|port=80-82| or \lstinline|port=[80-82]|), and how the specification of conditions and actions are started and terminated (e.g., to put a comma before processing the following capability). The example in Fig.~\ref{lst:concatenator} presents how to treat multiple values for IpTables source ports and how this affects the use of specific IpTables' extensions.

\begin{figure}[tb]
\begin{adjustbox}{max width = .49\textwidth}
\begin{lstlisting}[basicstyle=\scriptsize\ttfamily]
<bodyConcatenator>
  <operatorType>union</operatorType>
  <realConcatenator>,</realConcatenator>
  <concatenatorCondition>
    <preVariable>elementRange</preVariable>
    <postVariable>elementValue</postVariable>
  </concatenatorCondition>
  <newCommandName>
    <realCommandName>-m multiport --sports</realCommandName>
  </newCommandName>
  <newCommandName>
    <realCommandName>! -m multiport --sports</realCommandName>
    <commandNameCondition>
        <attributeName>operation</attributeName>
        <attributeValue>NOT_EQUAL_TO</attributeValue>
    </commandNameCondition>
  </newCommandName>
</bodyConcatenator>
\end{lstlisting}
\end{adjustbox}
\caption{Concatenator}
\label{lst:concatenator}
\end{figure}

Finally, the \class{CapabilityTranslationDetails} class describes the dependencies among capabilities, e.g., if a capability can only be present if other ones are already present or absent in the same rule, also allowing to set their values (\texttt{dependency}). 
For instance, in IpTables, the source and destination ports can only be used if a condition on the protocol type requiring UDP or TCP is in the same rule, as described in \changed{Fig.~\ref{lst:dependency}.}

\begin{figure}[t]
\begin{adjustbox}{max width = .49\textwidth}
\begin{lstlisting}[basicstyle=\scriptsize\ttfamily]
<dependency>
    <presenceOfCapability>IpProtocolTypeConditionCapability
    </presenceOfCapability>
    <presenceOfValue>tcp</presenceOfValue>
</dependency>
\end{lstlisting}
\end{adjustbox}
\caption{Dependency}
\label{lst:dependency}
\end{figure}

An example of an abstract policy for IpTables can be seen in Fig.~\ref{lst:iptables_rule}. The rule, added in the OUTPUT chain, drops TCP packets from a specific IP if received on a given network interface, while all the other packets are accepted. 
This policy has been validated against the 
\lstinline|language_ipTables.xsd| language, the automatically generated schema for IpTables, and translates to a file containing:
\begin{lstlisting}[basicstyle=\scriptsize\ttfamily]
iptables -A OUTPUT -i eth0 -p TCP -s 192.168.1.1 -j DROP
iptables -P OUTPUT ACCEPT 
\end{lstlisting}

\begin{figure}[tb]
\begin{adjustbox}{max width = .49\textwidth}
\begin{lstlisting}[basicstyle=\scriptsize\ttfamily]
<policy nsfName="IpTables" targetRuleSet="INPUT" 
    xmlns:xsi="http://www.w3.org/2001/XMLSchema-instance"
    xsi:noNamespaceSchemaLocation="language_ipTables.xsd">
<rule id="0">
    <externalData type="priority">1</externalData>
    <appendRuleActionCapability>
        <chain>OUTPUT</chain>
    </appendRuleActionCapability>
    <inputInterfaceActionCapability>
	<inFa>eth0</inFa>
    </inputInterfaceActionCapability>
    <ipProtocolTypeConditionCapability operator="exactMatch">
        <capabilityValue>
		<exactMatch>TCP</exactMatch>
	</capabilityValue>
    </ipProtocolTypeConditionCapability>
    <ipSourceAddressConditionCapability operator="exactMatch">
    <capabilityIpValue>
        <exactMatch>192.168.1.1</exactMatch>
    </capabilityIpValue>
    </ipSourceAddressConditionCapability>
    <rejectActionCapability/>
</rule>
<defaultActionCapabilitySpec>
    <acceptActionCapability/>
</defaultActionCapabilitySpec>
</policy>
\end{lstlisting}
\end{adjustbox}
\caption{An MLP for IpTables}
\label{lst:iptables_rule}
\end{figure}


\subsection{ResolutionStrategyDetails class}
\label{sec:ResolutionStrategyDetails}
\class{ResolutionStrategyDetails} class serves to characterise the external data that a Resolution Strategy uses so those values are properly generated during translation (e.g., \texttt{requiredExternalData}).

For example, the \textit{First Matching Rule} resolution strategy, \ie the default one for IpTables, associates rules to integers to express the priority. Based on priorities, the Translator will append rules to the selected chain in the proper order.

\subsection{Addressing other translation issues}

In most cases, but more frequently for conditions, there is the need to express more values when specifying a single capability. 
The way to specify multiple values varies with the type.
The simplest types, like ports or IP addresses, are usually specified individual values, ranges, or both (e.g., port=80,85-90,101,1024-).
More complex types, like URLs or application-level payloads, usually rely on regular expressions\footnote{Note that IP ranges are sometimes specified with a regex, but it is an overly sophisticated approach that easily maps to the set and range case.} 

The Abstract Languages generated from the \nsf instances have no limitations in the specification formats, \ie they support sets and ranges for integer-based types, sets and ranges (also in CIDR notation) for IP addresses, and Perl regex for string-based capabilities.

\changed{However, when generating the low-level configurations, some \nsf{}s may not fully support all the expressiveness allowed at the abstract level.
For example, prior to the development of the multiport extension, IpTables only accepted a single port number as a condition.
Theoretically, multiple values within a single capability are associated using a set union. Thus, when a union at the capability level is unsupported, it needs to be expanded into a union of individual rules.  The Cartesian Product needs to be used to expand the cases when multiple value capabilities are specified in a single rule.
The Translator understands that an expansion is needed by interpreting data from the corresponding \class{CapabilityTranslationDetails}. 
The attributes of the \class{CapabilityTranslationDetails} serve to customise this process. The \texttt{PreferredExpansionType} attribute explicitly states how the expansion must be performed.
By default, the Translator ensures a minimum number of rules are generated. In the worst case, e.g., \nsf{}s that only accept a single value for each condition, the number of generated rules could be large. Nonetheless,  we have not encountered these situations in all the \nsf{}s we have analyzed.
}

\section{Implementation}\label{sec:workflows}\label{sec:implementation}
\begin{figure*}
\centering
\begin{adjustbox}{max width = \textwidth}
\scalebox{1}{
\begin{tikzpicture} [node distance=1cm, font=\footnotesize]
   

\definecolor{verdino}{rgb}{0.698, 0.949, 0.733}
\definecolor{azzurrino}{rgb}{0.65, 0.85, 1}
\definecolor{giallino}{rgb}{1.0, 0.925, 0.6}

\node (SCMUML)[rectangle, rounded corners=5pt, text centered, draw=black, fill=red!20, align=center, inner sep=10pt,text width=3.5cm]{
    SCM UML
};

\node (XMIConv)[diamond, rounded corners, text centered, draw=black, fill=giallino!40, aspect=2.5, below=0.4cm of SCMUML, align=center, inner sep=5pt, text width=1.8cm]{
    XMI Converter
};

\draw [arrow, ->, line width=2pt] (SCMUML) -- (XMIConv);

\node (SCMXSD)[rectangle, rounded corners=5pt, draw=black, fill=gray!20, right=1cm of XMIConv, align=center, inner sep=5pt,text width=1.8cm]{
    SCM XSD
  };

\draw [arrow, ->, line width=2pt] (XMIConv) -- (SCMXSD);

\node (NSFCat) [rectangle, rounded corners=5pt, text centered, align=center, text width=3cm, right=.5cm of SCMXSD, yshift=1.6cm] {
    NSF Catalogue
};

\node (el3) [rectangle, text centered, align=center, fill=none, text width=0.3cm, below=0.01 cm of NSFCat, xshift=0.48cm, minimum height=1cm]{
    \textbf{...}
};

\node (el1) [rectangle, rounded corners=5pt, text centered, align=center, text width=0.5cm, left=0.02cm of el3, fill=white!20, minimum height=1cm, draw=black] {
    NSF\\ 2
};

\node (el2) [rectangle, rounded corners=5pt, text centered, align=center, text width=0.5cm, left=0.15cm of el1, fill=white!20, minimum height=1cm, draw=black] {
    NSF\\ 1
};
\node (el4) [rectangle, rounded corners=5pt, text centered, align=center, fill=none, text width=0.5cm, right=0.02cm of el3, fill=white!20, minimum height=1cm, draw=black]{
    NSF\\ n
};

\begin{scope}[on background layer]
\node(NSFCatBackground)[rounded corners=10pt, fill=verdino!40, fit=(NSFCat)(el2)(el4), draw=black, inner sep=5pt, anchor=center, text width=3.2cm] {};
\end{scope}

\draw [dashedarrow, ->, line width=2pt] ([yshift=0pt]NSFCatBackground) -| (SCMXSD);

\node (AbsLangGen)[diamond, rounded corners, text centered, draw=black, fill=giallino!40, aspect=3.5, right= of el4, align=center, inner sep=2pt,text width=2.3cm, yshift=10pt]{
    Abstract Language Generator
};

\draw [arrow, ->, line width=2pt] ([yshift=2.5pt ]NSFCatBackground.east) -- ([xshift=2pt ]AbsLangGen.west);

\node (NSFName)[rectangle, rounded corners=5pt, text centered, draw=black, fill=white!20, align=center, inner sep=5pt, text width=1.8cm, above right=-3pt and 1cm of AbsLangGen]{
    NSF Name
};

\draw [arrow, ->, line width=2pt] (NSFName.west) -- (AbsLangGen);

\node (CaM)[diamond, rounded corners, text centered, draw=black, fill=azzurrino!40, aspect=3.5, align=center, inner sep=2pt,text width=2.5cm, below=0.5cm of NSFCatBackground ]{
    Catalogue Manager
};

\node (Transl)[diamond, rounded corners, text centered, draw=black, fill=giallino!40, aspect=2.5, align=center, inner sep=2pt,text width=1.5cm] at ([yshift=+25pt]CaM -| AbsLangGen){
    Translator
};

\node (NSFAbsLang)[rectangle, rounded corners=5pt, text centered, draw=black, fill=white!20, align=center, inner sep=5pt, text width=1.8cm, right=1cm of Transl]{
    NSF Abstract\\Language
};

\draw [arrow, ->, line width=2pt] (AbsLangGen) -- (NSFAbsLang);

\node (LLC)[rectangle, rounded corners=5pt, text centered, draw=black, fill=white!20, align=center, inner sep=5pt, text width=1.8cm, below right=2cm and 1cm of AbsLangGen]{
    LLC
};

 \node (MLP)[rectangle, rounded corners=5pt, text centered, draw=black, fill=white!20, align=center, inner sep=5pt, text width=1.8cm] at (Transl |- LLC){
    MLP
};

\draw [arrow, ->, line width=2pt] (NSFAbsLang.west) -- (Transl);
\draw [arrow, ->, line width=2pt] (Transl) -- (LLC.west);
\draw [arrow, ->, line width=2pt] (MLP) -- (Transl);

\node (RefEng)[diamond, rounded corners, text centered, draw=black, fill=azzurrino!40, aspect=2.5, align=center, inner sep=2pt,text width=1.5cm, below=0.5cm of CaM ]{
    Refinement\\Engine
};

\node (RemMod)[diamond, rounded corners, text centered, draw=black, fill=azzurrino!40, aspect=2.5, align=center, inner sep=2pt,text width=1.6cm, left=1.5cm of RefEng ]{
    Remediation\\Module
};

\draw [arrow, ->, line width=2pt] (NSFCatBackground) -- (CaM);
\draw [dashedarrow, ->, line width=2pt] (RefEng) -- (CaM);
\draw [arrow, ->, line width=2pt] (RefEng.east) -| ++(0.8,1.275) -- (MLP.west);

\node (CTIRep)[rectangle, rounded corners=5pt, text centered, draw=black, fill=white!20, align=center, inner sep=5pt, text width=1.8cm, left=14pt of RemMod]{
    CTI Report
};

\node (MISPRep)[rectangle, rounded corners=5pt, text centered, draw=black, fill=white!20, align=center, inner sep=5pt, text width=1.8cm, below=0.8cm of CTIRep]{
    MISP Report
};

\node (Recipes)[rectangle, rounded corners=5pt, text centered, draw=black, fill=white!20, align=center, inner sep=5pt, text width=1.8cm, above=0.8cm of CTIRep]{
    Recipes
};

\node (Rems)[rectangle, rounded corners=5pt, text centered, draw=black, fill=white!20, align=center, inner sep=5pt, text width=1.8cm] at (RemMod.south |- MISPRep.center){
    Remediations
};

\node (HLP)[rectangle, rounded corners=5pt, text centered, draw=black, fill=white!20, align=center, inner sep=5pt, text width=1.8cm] at (RefEng.south |- MISPRep.center){
    HLP
};

\node (NetLand)[rectangle, rounded corners=5pt, text centered, draw=black, fill=white!20, align=center, inner sep=5pt, text width=1.8cm, anchor=south] at ($(Rems.south)!0.5!(HLP.south)$) {
    Network Landscape
};

\draw [arrow, ->, line width=2pt] (Recipes.east) -- (RemMod);
\draw [arrow, ->, line width=2pt] (CTIRep.east) -- (RemMod);
\draw [arrow, ->, line width=2pt] (RemMod) -- (MISPRep.east);
\draw [arrow, ->, line width=2pt] (RemMod.south) -- (Rems.north);
\draw [arrow, ->, line width=2pt] (NetLand.north) -- (RemMod);
\draw [arrow, ->, line width=2pt] (NetLand.north) -- (RefEng);
\draw [arrow, ->, line width=2pt] (HLP) -- (RefEng);
\draw [dashedarrow, ->, line width=2pt] (RemMod) -- (RefEng);
\draw [dashedarrow, ->, line width=2pt] (RemMod) |- (CaM);

\matrix (legend) [draw, rounded corners=5pt, above left, yshift=8pt, matrix of nodes, nodes in empty cells] at (current bounding box.south east) {
  \node(legendArt)[rectangle, rounded corners=2pt, draw=black, fill=white!20, align=center, inner sep=5pt, label=right:Artefact] {}; & 
  \node(legendArr1)[right=0cm of legendArt, xshift=-1cm]{};
  \node(legendArr2)[right= of legendArr1]{Input / Output};
  \draw [arrow, ->, line width=2pt ] (legendArr1) -- (legendArr2); \\
  \node(legendTool)[diamond, rounded corners=2pt, draw=black, fill=giallino!40, align=center, label=right:\scm Tool] {};& 
  \node(legendDash1)[right=0cm of legendTool, xshift=-1cm]{};
  \node(legendDash2)[text width=1.8cm, right= of legendDash1]{Component Interaction / Dependency};
  \draw [dashedarrow, ->, line width=2pt ] (legendDash1) -- (legendDash2);\\
   & \\
   \\
};

\node(legendComp)[fit=(legend-3-1)(legend-4-1)(legend-4-1),diamond, rounded corners=2pt, draw=black, fill=azzurrino!40, align=center, scale=0.4, label=right:Validation Component] {};

\end{tikzpicture}
}
\end{adjustbox}
\caption{Visual representation of the components built upon the \scm and their interactions.}
\label{fig:diagram}
\end{figure*}

We have implemented an entire ecosystem of components; some serve to manage the capabilities, and others use them to perform more sophisticated tasks, which are depicted in Fig.~\ref{fig:diagram}. 

The \acrlong{securitycapabilitymodel} is represented with the UML dialect of Modelio\footnote{\url{https://www.modelio.org}}, an open-source modelling environment for UML.
However, we needed XML technologies for practical usability and to be applied to our use cases.
The \textit{XMIConverter} tool automatically generates the XMLSchema (named \lstinline|capability_data_model.xsd|) describing the UML classes and properly mapping the original UML features. This file contains all the data types needed to model the security controls.

\acrlong{securitycontrol}s have been described in separate XML files and validated against capability XML Schema. These
XML files, rooted at the \class{NSF} class, list all the security capabilities, the associated policy details and all the translation instructions as \class{LanguageGenerationDetails} instances.
The Catalogue is an XML file that includes the \seccon{} files to use.

Given the name of an \nsf and the XML file describing it, the \textit{Abstract Language Generation} tool produces the XMLSchema file describing its abstract language that can be used to validate \mlp policies for that NSF, written in XML.
These two tools have been implemented in Java. They are expected to be executed occasionally, i.e., when changes are implemented in the model or new security controls are added to a catalogue.

The capability model can be managed and accessed through a web service named \cam{}, which exposes a user-selected NSF Catalogue through an XML DB. We have used BASE~X\footnote{\url{https://basex.org}} which provides various visualizations to explore data interactively and 
implements a RESTXQ service, enabling the development of web applications in XQuery.
A set of endpoints is exposed, as a set of XQueries, that implement essential functions/workflows:
\begin{itemize}
    \item \textit{Comparison}: it receives the names of two \nsf{} and, if these are present in the catalogue, checks if the sets of capabilities they own are equivalent, their intersection is empty, or one is contained in the other one. It returns the set of shared capabilities.
    
    \item \textit{Substitute}: it receives the name of an \nsf, then the service proposes other security controls in the catalogue that could be employed instead;
    
    \item \textit{Policy enforcement}: it receives an \mlp, and the service will identify other security controls that support the capabilities required to enforce the \mlp. Knowing this information, users can select one \nsf and use the Translator to obtain the policy in the specific language of that security control.
    
    \item \textit{\nsf search}: it receives a comma-separated list of capabilities and returns the list of \nsf{}s supporting all of~them. 
    
\end{itemize}

The \refeng is a component that refines requirements for securing a computer network, expressed as \hlp statements, into \mlp for the security controls in the target network.
\hlp statements are authorization policies expressed in the subject-action-verb-options paradigm \cite{nist-abac}.
\refeng analyses the subjects, objects, and optional fields present in the \hlp{}s and associates them with the entities in the target network; the \hlp{}s are then interpreted semantically to determine the capabilities needed for its enforcement. 
After identifying the required capabilities, the target network is explored. 
By querying the \cam{}, the \refeng identifies the \nsf{}s that own them, also proposing to add new controls from the Catalogue to remediate non-enforceable policies.

Once the usable controls are known, the selection of the ones to use is done according to a refinement strategy (e.g., defence in-depth, minimization of the security control to configure, minimize the delay). 
Finally, \refeng generates the \mlp{}s for all the selected \nsf{}s.

The refinement has been implemented as a forward reasoning task that semantically analyzes and enriches via inference using CLIPS\footnote{\url{https://clipsrules.net/}} and its Python bindings.
Its features can be accessed natively, as Python script or offered via a web service that exposes the following actions:
\begin{itemize}
    \item \emph{Upload}: allows the upload of the \hlp in XML format;
    \item \emph{Upload Landscape File}: allows to upload a file containing the target network description;
    \item \emph{Refinement}: perform an interactive refinement where the user is asked to provide missing information or make decisions when alternative refinement options are available (e.g., to select the devices to use); 
    \item \emph{Non-interactive Refinement}: the refinement process is performed using the user-selected refinement strategy;
    \item \emph{\mlp Generation}: generates the \mlp policies from the facts contained in the reasoning engine's knowledge base and then outputs the list of involved \nsf{}s;
    \changed{\item \emph{Download \mlp}: download the \mlp policies for one or all of the involved \nsf.}
\end{itemize}

The \textit{Translator} is the tool that, given an \mlp and the Abstract Language of the target security control, generates the configuration settings that map an input \mlp into the NSF-specific language.
This tool is available both as a web service and natively as a JAR.


Finally, the \remeng is the component that orchestrates the remediations to security incidents/events notified by Cyber Threat Intelligence (CTI) reports.
The \remeng interprets the reports (e.g., infers the type of threat and the impacted entities), identifies the applicable remediations, and evaluates their potential effectiveness.

Remediations are enforced through the procedural execution of a set of actions, called Recipes, expressed in an intent language. \remeng recipes are interoperable with the CACAO standard and follow the best practices of playbook-based remediation description \cite{threatremediationset}.
These recipes consist of high-level instructions that may propose to add new rules to existing \seccon (e.g., new firewall filters) by proposing new \hlp{}s, suggest network changes (e.g., isolating or shutting down machines), or propose adding new \seccon{}s to mitigate the threats. The \cam helps identify the controls to use from the ones available in the NSF Catalogue.

The \remeng can automatically select the most effective recipe, as security experts rated every recipe's effectiveness on individual incidents. Otherwise, users can manually choose a suitable one from a GUI.
Based on a DSL framework employing a visitor pattern and a PEG grammar, an interpreter gathers all necessary data for recipe deployment and the execution of specific actions, and then runs the recipe stepwise.
The \remeng interfaces with the \refeng to trigger the refinement process of \hlp statements that may be mandated by the recipe or with the network infrastructure to initiate changes to the network or routing.

\changed{
\subsection{Security and complexity considerations}
It is important to emphasize that the \scm, the artefacts it generates (e.g., the XML Schemas), and the associated catalogues are all sensitive assets. The effectiveness of security enforcement and remediation relies heavily on their content. Modifying these files can cause refinement systems or remediation tools to fail or produce errors. 
As a result, these assets must be protected against attacks aimed at altering them, whether those attacks come from external sources or insiders. Protecting these assets is not different from safeguarding other pieces of security-sensitive information. 
Therefore, these assets must be protected well against attacks aiming at modifying them both from externals and insiders. 
Protecting these assets is not different from protecting other security-sensitive information.
We expect only the vendors or the teams developing open-source security controls may need to modify the model or the catalogues. 
Hence, the data related to the model could be authenticated using digital signatures, thereby significantly reducing the risks associated with most insider attacks.

In this scenario, the complexity related to the management of the \scm and its ecosystem would be largely compensated by the advantages given by the automated (or semi-automated) security policy enforcement and remediation that can be built on top of our model.
}

\section{Validation}\label{sec:validation}



\changed{Our validation focused on the three scenarios in Section~\ref{sec:motivating}. 
The reference network (available in our repository\footnote{\url{https://github.com/aldobas/SecurityCapabilityModelPub}}) served as a representative example of a medium-sized company network, with ten subnets interconnected by eight security-enabled nodes, each implementing one or more security controls.}

\subsection{Cybersecurity market and Vendor lock-in}\label{sec:validation:uc1}

\begin{table}[tb]
    \centering


    \begin{tabularx}{.49\textwidth}{l}
        \toprule
        \multicolumn{1}{c}{\textbf{Capabilities of the PF-PacketFilter}}\\
        \midrule
        AcceptActionCapability, RejectActionCapability,\\
        TcpFlagsConditionCapability, DestinationPortConditionCapability,\\
        SourcePortConditionCapability,  IpDestinationAddressConditionCapability,\\
        IpSourceAddressConditionCapability, IpProtocolTypeConditionCapability,\\
        InterfaceConditionCapability, 
        NumberConnectionsConditionCapability,\\
        LimitSAddrConditionCapability, \textit{StateInterfaceBoundConditionCapability},\\
        \textit{OutputInterfaceConditionCapability}, \textit{InputInterfaceConditionCapability},\\
        \textit{MaxRateConnectionsConditionCapability}\\
        \toprule
        \end{tabularx}
    \label{tab:test}
    \caption{The PF capabilities (PF-specific ones are in italic).}
\end{table}

The validation of the first proposed use-case simulated the migration of the company's internal firewall from  IpTables to a commercial product based on the pfsense solution\footnote{\url{https://www.pfsense.org/}}, which builds on the  BSD Packet Filter (PF).
This validation aimed to assess the migration's feasibility, mimicking what security operators would need to perform in a hypothetical company with this scenario.

Two members of our research group assumed the role of security operators. They were asked to describe PF according to the \scm, reusing, whenever possible, what was already available in the Data Model and Catalogue files for IpTables.
Indeed, PF provides capabilities akin to IpTables but exhibits its own peculiarities, as commonly observed in such devices.

As the initial catalogue did not model PF, the assessment was preceded by a meticulous analysis of the PF features and syntax. 
After this analysis and a model adaptation phase, the features of PF were successfully integrated into our \scm.
The resulting set of capabilities from modelling PF according to the capability model schemas is listed in Table~\ref{tab:test}. 

We had to support in the model the case of IpTables that only allows for the specification of conditions on the network interface name without considering the direction (i.e., in, out, forward), as this concept is implicitly specified using the target chain name. On the other hand, PF puts both name and direction in the same condition. 
To represent this case, separate conditions capabilities were introduced.

Significant challenges surfaced during the modelling of PF stateful capabilities.
PF misses some state filtering capabilities available to IpTables, e.g., the independent management of TCP connections based on the knowledge of TCP three-way handshake protocol states (e.g., NEW, ESTABLISHED, and INVALID connections) typical behaviours (Request/Reply), or IP management like ICMP packets (e.g., RELATED). 
In PF, all the packets related to a TCP connection allowed by an explicit rule are allowed as well by default. Otherwise, the stateful behaviour can be fully disabled (\lstinline|no state|). Finer-grained conditions on the TCP flags are managed as stateless conditions.
However, some state-related features available in PF are lacking in IpTables, e.g., PF supports run-time options that determine if packets belonging to an allowed connection can arrive at any interface and direction (\lstinline|floating|) on the same interface as the first rule match (\lstinline|if-bound|).

The two packet filters also exhibit differences in managing conditions on packet counters and connection bandwidth, as PF only allows basic rate-limiting capabilities on the number of established connections (\lstinline|max-src-conn| and \lstinline|max-src-conn-rate|).

Therefore, to support both packet filters, the \scm introduced finer-grained condition capabilities than the ones only needed for IpTables to better depict this scenario.

To support translation, we added new \class{TranslationCapabilityDetails} instances and adjusted the Translator to build one single low-level condition from data taken from more than one capability at \mlp-level (e.g., \lstinline|max-src-conn| which merges data from the \class{NumberConnectionsConditionCapability} and \class{LimitSAddrConditionCapability} and add only once the \lstinline|keep state| keyword before introducing stateful options). 

\changed{This simulation demonstrates the model's flexibility in integrating new security controls. Supporting the general case of stateful packet filters may require additional refinements and adaptations to enhance stateful capabilities, which we will address as future work. We recognize that incorporating all security controls can be time-consuming. However, we anticipate that uncommon cases, particularly those as challenging as managing stateful conditions, will become increasingly rare.
}
%




After PF was added to the Catalogue, all the queries in Section~\ref{sec:implementation} were available and confirmed that PF capabilities intersect with the ones offered by IpTables.
Moreover, with the \textit{Policy enforcement} query, we specified an \mlp for a generic packet filter that was recognized as enforceable by both IpTables and PF. 
That MLP was readily translated for both devices. 
Moreover, we translated IpTables policies into PF policies using the Translator\footnote{Note that the conversion from IpTables configurations to \mlp has been made with an \textit{ad hoc} script. 
The Translator does not provide generic features for low- to medium-level transformation. 
Obtaining \mlp from \llc is feasible and was considered initially, but has limited research applications and is more tedious to implement because languages allow different methods for expressing the same policies.}. 
This underscores how PF may serve as a viable replacement for IpTables.

These experiments proved that the \scm can be used to evaluate how different security controls can handle high-level requirements and select the most suitable ones, hence proving our research objective $\mathrm{RO}_1$ presented in Section~\ref{subsec:motivating-uc1}.
\changed{In concrete scenarios, our model can be used in workflows that allow policies to be translated between different formats, which renders vendor lock-in more complex to achieve, hence proving our research objective $\mathrm{RO}_2$.
Moreover, it showed enough flexibility to cope with adding more \acrlong{securitycontrol}s.}


\subsection{Policy refinement}\label{sec:validation:uc2}

\changed{The refinement has been initially validated against an abstract network inspired by a real-world scenario.}
The network comprises ten different subnets, interconnected by six security controls, namely four filtering devices (IpTables) and two VPN gateways (XFRM and Strongswan).
In addition, three general-purpose entities have been introduced as actors for the policy refinement process: two legitimate network users and an external network segment representing malicious users.


Authorization and channel protection policies were expressed using \hlp{} statements.
In the first step, the \refeng identifies the capabilities needed to implement the policies. 
Then, it looks for the security controls in the catalogue owning these capabilities, and it searches for the availability of these controls in the path between subjects and objects. 
When multiple controls were available to implement a policy, the user was asked to select the ones to use via an \textit{ad hoc} web GUI (see Table~\ref{tab:controls}. 
We manually verified that the options offered were indeed correct: all the devices provided the required capabilities and belonged to the correct network paths. An optimization model could help this selection process; it will be investigated in future work.
Once the \nsf{}s were selected, the \refeng generated the \mlp{}s by calling the Translator. 
A manual assessment proved that the generated \mlp{}s and \llc correctly implemented the input \hlp{}s and the \llc were properly imported by the target \seccon{}s.

\begin{table*}[]
\label{tab:controls}
\caption{Sample \hlp statements and \seccon available to enforce them in the target network.}
\centering
\begin{tabularx}{.9\textwidth}{lllX}
\toprule
\multicolumn{1}{l}{\textbf{Subject}} & \multicolumn{1}{l}{\textbf{Action}} & \multicolumn{1}{l}{\textbf{Object}} & \multicolumn{1}{l}{\textbf{Available Security Control}} \\ \midrule
Bob                                    & is authorized to access              & internet traffic                     & Path 1: firewall-1 (IpTables),\\& & & Path 2:  firewall-1 (IpTables), firewall-2 (IpTables)                                            \\ \hline
Alice                                  & protect integrity                    & Bob                                  & Path 1: firewall-1 (XFRM), vpn-gateway-1 (StrongSwan)                      \\ \hline
Malicious\_User                        & is not authorized to access          & Alice                                & Path 1: firewall-1 (IpTables), firewall-2 (IpTables)\\& & & Path 2: firewall-2 (IpTables) \\
\bottomrule
\end{tabularx}
\vspace{.2em}

\end{table*}


After the internal assessment, the \refeng was used by the Consortium of the EC-funded FISHY project. It was integrated into the FISHY architecture and used in the project's Use Cases \cite{d4.4_fishy,d6.4_fishy}. 
Supporting the Use Case scenarios required the addition of two more security controls: 
\begin{itemize}
    \item an authorization control for an Ethereum app to ban users misbehaving based on their Walled ID and Distributed ID;
    \item a generic device created to export firewall configurations in the format useful to interact with people who had to load configurations manually in a SaaS scenario.
\end{itemize}   

In this phase, the validation was performed at the Consortium level. 
The \refeng was used to refine high-level security requirements for protecting the Use Case networks. The Translator was then used to obtain the \llc, which were deployed into the target \seccon of the Use Case networks. 
By means of probe packets and tentatives of misusing the network, the Consortium validated that the security policies were correctly enforced\footnote{https://www.youtube.com/channel/UCSDpfCPvFNjRS3RemG0iNQQ}.


The validation proved that the \refeng, built around the \scm, proves that the research objectives established in Section~\ref{subsec:motivating-uc2} can be reached. The engine automatically identifies the security controls required for the specified policies ($\mathrm{RO}_3$) and enables the refinement from \hlp{}s to \llc{}s ($\mathrm{RO}_4$).
\changed{
Moreover, the \hlp{}s are based on the application-specific attack model and are independent of the \scm framework, whose purpose is to enable automated refinement via abstraction, which ensures adaption to different scenarios and contexts.
}

\subsection{Incident response}\label{sec:validation:uc3}

The validation of this scenario required the implementation of an incident response pipeline to protect a software network environment. 
Reactions involved two types of remediations. 
We could change the network topology through a software network management layer, \eg adding new security controls or redirecting traffic flows.
Moreover, reactions involved the reconfiguration of security controls through the \refeng.

\changed{In the incident response context, an important requirement for enforcing the playbook abstraction in security procedures is the need to operationalize workflows. These workflows are represented by high-level security actions such as ``block'' or ``audit''. This requires selecting a suitable device to enforce the specified action at runtime. Once identified, a corresponding low-level configuration must be generated for the target operational environment to implement the policies effectively.
In this sense, the \scm concretely demonstrates its effectiveness firstly, by facilitating the selection of security mechanisms capable of enforcing the specified actions, and secondly, by enabling the generation of concrete policies and low-level configurations necessary for the playbook to take effect~\cite{risposta-incidentresponse}.}

The effectiveness of the incident response was tested within the distributed systems of the use cases of the PALANTIR\footnote{\url{https://www.palantir-project.eu/}} and FISHY\footnote{\url{https://fishy-project.eu/}} projects.
Upon detection of a new threat, the system was designed to support automatic selection of remediations  (in PALANTIR) or manual selection from a list of alternatives (in FISHY). \changed{In both cases, the remediations were dictated by the underlined scenarios' attack models, and the \scm provided the abstraction layer.}
In PALANTIR, the \remeng was responding to a crypto-mining 
and DDoS botnet threats. The recipes for remediating these scenarios required a security control capable of traffic filtering at layer 4 to isolate the infected machines.
In FISHY, the \remeng was reacting to increasing authentication failures to a distributed ledger based on Ethereum. The recipes proposed to filter traffic at layer 7 (URL) and add banning rules for specific Distributed IDs and Wallet IDs to the ledger authorization module.
Moreover, the reactions also remediated DDoS and brute force attacks. In this case, the recipes proposed filtering traffic at layer 7 or 4.
In all cases, when a needed \seccon{} was unavailable in the software network, the recipes asked for the deployment of a capable one in the correct portion of the network to protect target hosts if needed. 
Finally, the security controls were properly reconfigured through the \refeng and the Translator.

During the design phase, stakeholders verified that recipes effectively mitigate the identified risks.
During the validation, the project Consortia used the tools to deploy new configurations. Then, they checked with experiments that the threats were actually mitigated by verifying that the attacks were rendered ineffective. 

In PALANTIR, in addition, the Remediation and threat-related information were shared as MISP reports, including the Recipe and metadata according to a structured template, including security capabilities, actions, deployment parameters, and enabling constraints defining the parameters necessary for correct execution \cite{threatremediationset}.

The practical validation conducted through the two projects leveraging the \scm confirmed fulfilling the requirements in Section~\ref{subsec:motivating-uc3}. Specifically, it demonstrates the feasibility of integrating an \scm-based workflow to ensure interoperability of various security controls into automated, playbook-based incident handling ($\mathrm{RO}_5$).
Moreover, this validation further ensures enhanced cyber threat intelligence sharing, leading to improved actionability of such information and providing analysts with deeper insights into threats and their effective mitigation or remediation strategies ($\mathrm{RO}_5$).

\section{Related works}\label{sec:related}

In earlier work, Basile et al.\ presented an ontology-based approach to abstractly translating policies into low-level configurations~\cite{basile2010ontology}. This approach exploited ontology as a tool for formal modelling knowledge about security controls. The representation of \seccon{} features was very coarse-grained due to the severe limitations posed by ontology-based reasoning and, hence, of limited usability. 
This work evolved into a more coherent framework for supporting refinement in software networks~\cite{basile2019ton}. Still, instead of modelling individual capabilities, this work used \seccon categories to implicitly determine a standard set of features they exposed. Hence, it was not able to capture the differences between \seccon{}s in the same category.

When speaking of modelling individual capabilities offered by \nsf{}s, the major efforts are presented by \emph{CapIM}~\cite{ietf-i2nsf-capability-05}, the information model developed by the I2NSF WG to provide a standard interface for control and management of \nsf{}s. The work described in this paper is a fork of \emph{CapIM}, which inherits the basic design principles and most of the Information Model. However, the data models and the features for translation are a clear improvement over the original \emph{CapIM}.

As shown by Geismann et al.~\cite{modeldriven-literature-review}, model-driven approaches towards security are currently a relevant research area, with various examples applied to cyber-physical systems. The reason is that having a formal model may eliminate threats and possible vulnerabilities in the early phases of the design while also providing flexibility during implementation. Similarly, adopting the same approach in the context of security functions results in a generalized representation of the entities that can be easily specialized to match the required security controls during implementation.

The D3FEND framework is another example of how helpful formalization is in cybersecurity \cite{kaloroumakis2021toward}. 
The D3FEND model focuses on a high-level semantic characterization of various countermeasures provided by cybersecurity products.
The D3FEND semantic model maps attacks, defence strategies or techniques, and digital artefacts, defined as generic computer systems or information used by them. 
The knowledge graph that the model provides enables practitioners to analyze and compare products based on their functionalities rather than relying on marketing materials. Moreover, this information can be used to assess the defence posture of a computer network infrastructure based on the countermeasures available.
As it focuses on abstract countermeasures, D3FEND works at a higher level than our \scm. Hence, it can be considered a complementary approach that can interoperate.

Analysing past research in automatic policy refinement and incident response is also interesting.
Despite being focused on firewalls, the work by Kovačević et al. offers the most recent systematic review of the literature related to automated policy refinement and low-level translation, highlighting a significant amount of work existing in the field~\cite{systematic-review-automatic-translation}. In this context, Rivera et al. introduced a two-level mechanism for policy translation and automatic incident response~\cite{rivera2019automatic}. However, the focus of said work is the development of a high-level model for a policy language and does not include a formal model that represents the security capabilities that must enforce such policies. 
Cheminod et al. developed an approach to refine and verify access control policies \cite{cheminod2019comprehensive}; the set of actions taken into consideration is limited and uncoupled from the security controls that should enforce them, an issue that can be addressed with the introduction of the \scm.

Leivadeas et al., in their survey~\cite{surveyonibn}, explore various facets of the intent-based networking (IBN) paradigm. They introduce five distinct concepts that characterize IBN, namely intent profiling, translation, resolution, activation, and assurance. In particular, translation enumerates the various approaches for converting intents into granular network configurations suitable for network devices, 
and activation addresses the mechanisms involved in deploying intents within the network system. 
Among the open challenges, they mention Intent fulfilment, which concerns the translation of intents into network policies and configurations. It is observed that translation mechanisms often involve multiple levels of refinement, each one extracting different details of the intent, ensuring consistency with intent provisions. Another crucial point is the need for vendor-agnostic fulfilment of intents, ensuring that infrastructure configuration mechanisms deriving from intents are effective across multi-vendor environments. 
Our \scm in an important step towards solving these issues.

\section{Conclusions}\label{sec:conclusions}
This paper presented the \acrlong{securitycapabilitymodel}, a precise and extensible formal model that describes several types of \seccon{}s. Its Information Model captures the general concepts, and its Data Model reports the capabilities needed to model packet filters and solutions for establishing secure channels.
Thanks to a model-driven approach, it allows for implementing methods for translating abstract policies into \seccon-dependent configurations. It also supports the automatic creation of abstract languages to describe vendor-specific policies.
The model proved effective in automating various crucial security tasks. Its basic features (e.g., comparing capabilities and identifying security controls based on the features required for enforcing policies) helped automate administrators' typical tasks, like migrating to new security controls.
These basic features allowed for building more advanced services, i.e., security policy refinement and handling security incidents with procedural remediations. 
All the components developed upon the model have been validated on realistic scenarios and real-world applications in the context of EC-funded projects, leading to extremely promising results.

While writing this paper, we are already adding a new category of \seccon, namely ModSecurity, an open source, cross-platform web application firewall (WAF) engine that can be configured with an event-based programming language that is more sophisticated than the ones we currently support. Moreover, we plan to add several open-source and commercial packet filters to our catalogue to capture precisely all the facets of these stateful devices.

%
From the practical point of view, we are investigating how to integrate our model with the MITRE D3FEND framework to propose contributions. \changed{
Moreover, the \scm may also be used alongside different security standards and guidelines, e.g., the ISO 27000 family and NIST SP~800-53, to help link conceptual frameworks, attack models, and practical tools concerning security controls. Moreover, we will further consider how our model may impact OT security, as per the ISA/IEC 62443 standard.
}
From the research point of view, we are investigating how to build an optimization model for determining the best resolution strategies in the \refeng and a risk-based evaluation of the mitigations proposed by the \remeng.
Furthermore, we are exploring the possibility of using the model for Retrieval-Augmented Generation to improve LLM performance in this policy-based management field. Moreover, as a long-term direction, we would like to use the model with AI techniques to develop methods that can autonomously build (the best) reactions to incidents.



\ifCLASSOPTIONcaptionsoff
  \newpage
\fi

\bibliographystyle{IEEEtran}
\bibliography{IEEEabrv,biblio}


\end{document}